\documentclass{article}[12pt]
\usepackage{epsfig}

\begin{document}

\title{\Large{\bf New Challenges in Cellular Automata  Due to Network Geometry --- Ferromagnetic Transition  Study}}
\vspace{9.0mm}
\author{\textsc{Danuta Makowiec }\thanks{Corresponding Author E-mail: fizdm@univ.gda.pl}\\
       {\small\it Institute of Theoretical Physics and Astrophysics, Gda\'nsk University} \\ 
       {\small \it ul.Wita Stwosza 57, 80-952~Gda\'nsk, Poland} \\
}
\maketitle

\begin{abstract}
{Self-organization to ferromagnetic  phase transition  in  cellular automata of spins governed by stochastic majority rule and when topology between spins is changed, is investigated numerically.  Three types of edge rewiring  are considered. The algorithm of  Watts and Strogatz  is applied at each time step to establish a network evolving stochastically (the first network). The preference functions are defined to amplify the role of strongly connected vertices (the second network). Demand  to preserve graph connectivity provides the third way of edge rewiring. Each of these processes  yields the different network:  small-world, scattered nodes with one strongly connected component, and scale-free network when rewiring is properly adjusted. The stochastic majority rule applied to these networks can lead or not to ferromagnetic transition. Classical  mean-field transition in case of the first and third network is observed. Scale-free ferromagnetic transition can be observed when rewiring is properly adjusted. }

\end{abstract}

\section{Introduction}
From the cellular automata origin - John von Neumman idea, when studying cellular automata what we have in mind there are ways in which cellular automata can be used as a tool for understanding the nature. By examining the similarities between human and technological systems and their organizations the importance of the network effect has been emerged, see \cite{BA99,BAReview} for review and bibliography. The lesson is that not only design of interactions matter greatly in reconstructing nature phenomena but the structure of connections. In particular,  asymmetry and heterogeneity matter significantly. Therefore the world of  cellular automata systems  should include the possibility to change the geometry on the underlying network. The following paper is a proposition in this direction.

For many physical applications, when considering the phenomena where many elements  interact locally, (i.e., only nearest neighbors interact with each other) then  it is enough to arrange the elements  in the regular way. The basic example is the model of localized spins, called Ising model --- a toy model for ferromagnetic phase transition study \cite{Ising}. The elements  are located in vertices of some regular lattice and the interactions are  set on static neighborhoods. Many cellular automata rules have been designed to simplify (though still represent) the  Ising interactions. The stochastic majority rule is the standard tool to imitate ferromagnetic interactions in cellular automata  because this rule is  linear approximation to interactions of the  Ising model,  see \cite{CAferro}. Lately, the critical phenomena observed in cellular automata with stochastic majority rule have been discovered as useful for investigations aimed on understanding dynamical properties of the neuropil --- the densely interconnected neural tissue structure of the cortex \cite{Kozma}.

Erdos and Reney proposed a model of random  arrangement of elements. The network is built of N vertices which are connected by an edge with probability $p$. Most features of the resulting network are determined by $p$, e.g., the average degree of a vertex $<k> = pN$ or the mean path between nodes. The degree  distribution is  Poisson distribution what indicates that most nodes have approximately the same number of links --- close to average. The mean path length between any two vertices  is proportional to the logarithm of the network size.  This property  is known as the small-world property and was discovered when networks of social relations were investigated  \cite{Milgram}. The investigations of many real structures, from  natural to  artificial origin,  ensure  that real networks are much different from the random network. They are highly non-uniform what effects  that the degree distribution follows the power-law decay rather than the exponential decay. 

The algorithm which transfers a regular lattice into so-called small-world was proposed by Watts and Strogatz \cite{WS98} and is based on rewiring with some probability $p$ edges of a regular lattice. However, the distribution of vertex degree in the Watts-Strogatz network is Gaussian. The Barabasi -Albert recipe for constructing a network with the power-law distribution of vertex degree requires two ingredients: the growth of the network and the preferential  attachment of new vertices added to a network at each time step \cite{BA99}. 

The Ising model  has been studied both analytically and in simulations in the small-world topology, \cite{Gitterman,Herrero} and in the free-scale topology \cite{Aleksiejuk}. 
In both cases a change of properties from the regular lattice to the mean-field characteristics is observed. 
Gitterman \cite{Gitterman} showed   analytically  that in the  small-world lattice the ferromagnetic ordered phase exists and the critical temperature depends on parameter $p$ of the  small-world topology. A  different behavior is observed in scale-free network. Here the infinite critical temperature  was found in the thermodynamic limit of infinite system, \cite{Aleksiejuk}.

In the following paper we investigate numerically how manipulations with the underlying regular lattice (rewiring and preferential attachment) influence  the network structure and  ferromagnetic properties of cellular automata of spins. The changes considered preserve both the number of nodes and the number of edges but the network is evolving since the connections between nodes are modified each time step. In Section 2 we present the model of cellular automata on evolving network. Then, in Section 3, we report results of simulation experiments aimed on testing critical properties in the system. Section 4 concludes our observations and states  questions for further investigations.

\section{Cellular automata on evolving network}

Let us consider typical cellular automata with ferromagnetic interactions. It means that the basic elements, called spins $\sigma_i$,  are  located in nodes of the square lattice of linear size L ($i\in 1,2,\dots,L^2$) and each spin is in one of the two following states: $up$  ($\sigma_i=1$) or $down$ ($\sigma_i=-1$). Each edge is named by two numbers which represent nodes connected by the edge.

The evolution step is two-part:
\begin{enumerate}
\item[(I)]--- rewiring: each edge $(i,k)$ sequentially  is rewired with  probability $p$:
\begin{enumerate}
  \item[(a)] independently of the node properties;
   \item[(b)] modified by preference to:
   \begin{enumerate}
     	\item[(i)]unlink from a node $k$th that is weakly connected to the graph:
     	 \begin{equation}
     	 p(k) = 1-\frac{deg(k)}{T}
     	 \end{equation}
     	 $deg(k)$ means degree of $k$th node, $ T$ - threshold for the preference;
     	\item[(ii)]link to a node $j$th  that is highly connected to the graph:
     	 \begin{equation}
     	      	 p(j) = \frac{deg(j)}{T}
     	 \end{equation}
   \end{enumerate}  	 
   \item[(c)] the preference (i) of (b) is restricted by the demand to preserve connectivity of the graph.
\end{enumerate}
\item[(II)] --- ferromagnetic interaction: the future state of a spin $\sigma_i$ is determined by
 the stochastic majority rule. The rule operates on  a present state of spins from the nearest-neighbor set $N_i(t)$.  Namely, with the probability $1-\varepsilon$:
\begin{equation}
\sigma_i(t+1)=\cases{
               sign(\Sigma_{j\in N_i(t)}\sigma_j(t)) &if $\Sigma_{j\in N_i(t)}\sigma_j(t) \neq 0$ \cr
               \sigma_i(t)   &otherwise \cr }
 \label{step}
\end{equation}
and with probability $\varepsilon$ the opposite to (\ref{step}) state is assigned to $\sigma_i(t+1)$
(a rule makes mistakes with probability $\varepsilon$) The ordinary interpretation of  a stochastic noise is that the noise is  the  invert of temperature.

\end{enumerate}
By this rule  the nearest-neighborhoods of nodes  are modified in each step asynchronously while   the majority rule is  applied synchronously.

It is known that the stochastic  majority rule  applied to the regular square lattice self-organizes the system to mimic  the ferromagnetic phase transition. Ferromagnetic features of stationary states depend critically on the value of $\varepsilon$ \cite{CAferro}.  These critical properties  are examined by observing features of the two following functions:
\begin{itemize}
 \item {\it magnetization} --- $$ m(t)= \frac{1}{L^2} \Sigma_{i=1}^{L^2} \sigma_i(t)$$ 
 \item{\it susceptibility }--- the variance of magnetization.
\end{itemize}

The magnetization is expected to change rapidly when the temperature noise $\varepsilon$ is growing, from non-zero magnetization state to zero-magnetization state. When passing the point of change   at $\varepsilon_{crit}$ the variance of magnetization becomes singular in the thermodynamic  limit of infinite system. In the finite system  the point of the critical change is depicted as the narrow maximum of susceptibility. The large value of susceptibility denotes that fluctuations of magnetization (homogeneous islands of one spin state) of any size are possible. 

By switching on the process of  edges rewiring,  a regular lattice  becomes a lattice with long range interactions. Bridges between distant nodes are established. The short-cut links, in general, break isolation of homogeneous islands that are present on the lattice due to  work of the majority rule. The described process  should effect both the value of magnetization in the stationary state and the point of critical change.

The simulation experiment goes as follows: starting with all spins $up$  state the two-part time step  for the fixed probability of rewiring $p$, fixed procedure of rewiring: (a), (b) or (c) and fixed value  of noise $\varepsilon$ is applied plenty of times to achieve the stationary state. For the  stationary state the  magnetization and susceptibility are collected. The properties of the network are studied by the distribution of vertex degree.

If the majority rule is considered at $\varepsilon=0$ (deterministic rule) then properties of stationary state depends critically on the magnetization of the initial state,  see \cite{ACRI2004}. It has been found that independently of L:
\begin{itemize}
\item     if $m(0) < 0$ then  the final state is {\it all spins $down$} configuration
\item     if $m(0) > 0$ then  the final state is {\it all spins $up $} configuration 
\end{itemize}
Therefore one can say that the cellular automata perform, so-called, {\it density task} with high accuracy.

\section{Ferromagnetic phase transition}
Starting with all spin UP on the lattice with linear size $L=200$ for each model of preferences, we apply the two-step dynamics: rewire an each edge with probability $p$ and  apply the stochastic majority  rule (\ref{step}) for nearest-neighbors with noise $\varepsilon$. After the stabilization time (here 5000 steps) we calculate  magnetization and susceptibility of the stationary state during the next 5000 steps. 

\subsection{Stochastic evolution of edges}
\begin{figure} 
    \begin{center} 
      \includegraphics[width=0.50\textwidth]{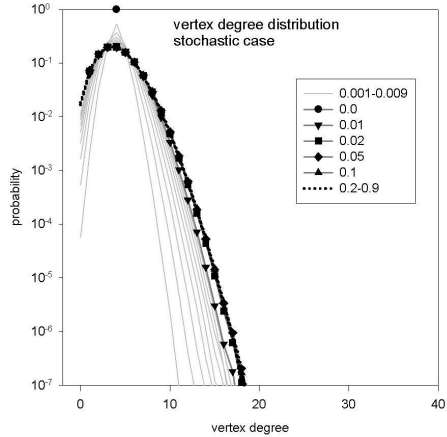}
       \end{center} 
\caption{\newline\small
Distribution of vertex degree in case of random rewiring for different $p$. } 
\label{stochastic}
\end{figure}

Gitterman \cite{Gitterman} showed   analytically  that in the  small-world lattice, obtained from a 1-dimensional ring where  edges are rewired with probability $ p$ and  $2k$ nearest neighbors interact according to the Ising hamiltonian, the crossover temperature to the mean-field critical behavior varies for $p<<1$ as $T(p) \propto k(k+1)/\ln(p)$, whereas the critical temperature scales as $T(p) \propto -2k/\ln(p)$. So that a ferromagnetic ordered phase exists for any finite $p$.

A typical trajectory of the magnetization obtained for pure stochastic rewiring, at any $p$, shows the rapid change in the value of stationary magnetization when the temperature noise $\varepsilon$ grows. The plot in Figure \ref{stochastic} presents the characteristic of the network connections. The  distribution of vertex degree is plotted on the log-scale to amplify its parabola-like shape. It means, that  the exponential network is obtained due to this  evolution. One should notice that  the distribution is almost independent of $p$, namely,  for  $p> 0.1$ all curves collapse into the same distribution.
\begin{figure} 
    \begin{center} 
      \includegraphics[width=0.99\textwidth]{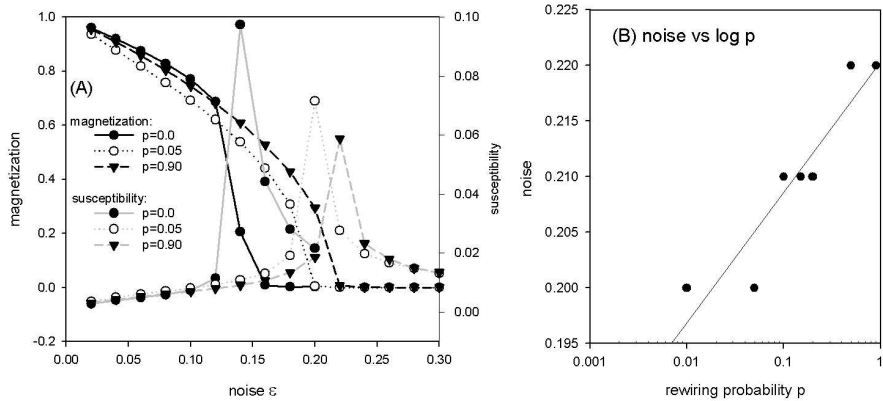} 
    \end{center} 
\caption{\newline\small
The points of phase transition $\varepsilon_{crit}$ for different  $p$ determined by the maximum in susceptibility (left panel) and  $\varepsilon_{crit}$ dependence on $p$ (right panel). The  stochastic rewiring case.} 
\label{epsi_crit}
\end{figure}

The point of the  critical change $\varepsilon_{crit}$  moves, see Figure \ref{epsi_crit}. 
If there is no rewiring (the classic cellular automata)  then the transition goes at about $\varepsilon_{crit}= 0.14$. If the lattice is rewired stochastically each time step then the transition point moves  to  $0.20 \pm 0.01$ at  $p=0.05$, and  to $0.22$ at $p=0.90$, see Figure \ref{epsi_crit}(left panel). Hence with increasing rewiring homogeneous islands are more stable against the temperature noise. Moreover, the  value of $\varepsilon_{crit}$ plotted against the log of rewiring probability $p$ scales linearly (Figure \ref{epsi_crit}(right panel)), what is   in the full agreement with the property found in the Ising model on the small-world network \cite{Gitterman,Herrero}.

\subsection{Evolution of edges with preferences}
\begin{figure} 
    \begin{center} 
    \includegraphics[width=0.90\textwidth]{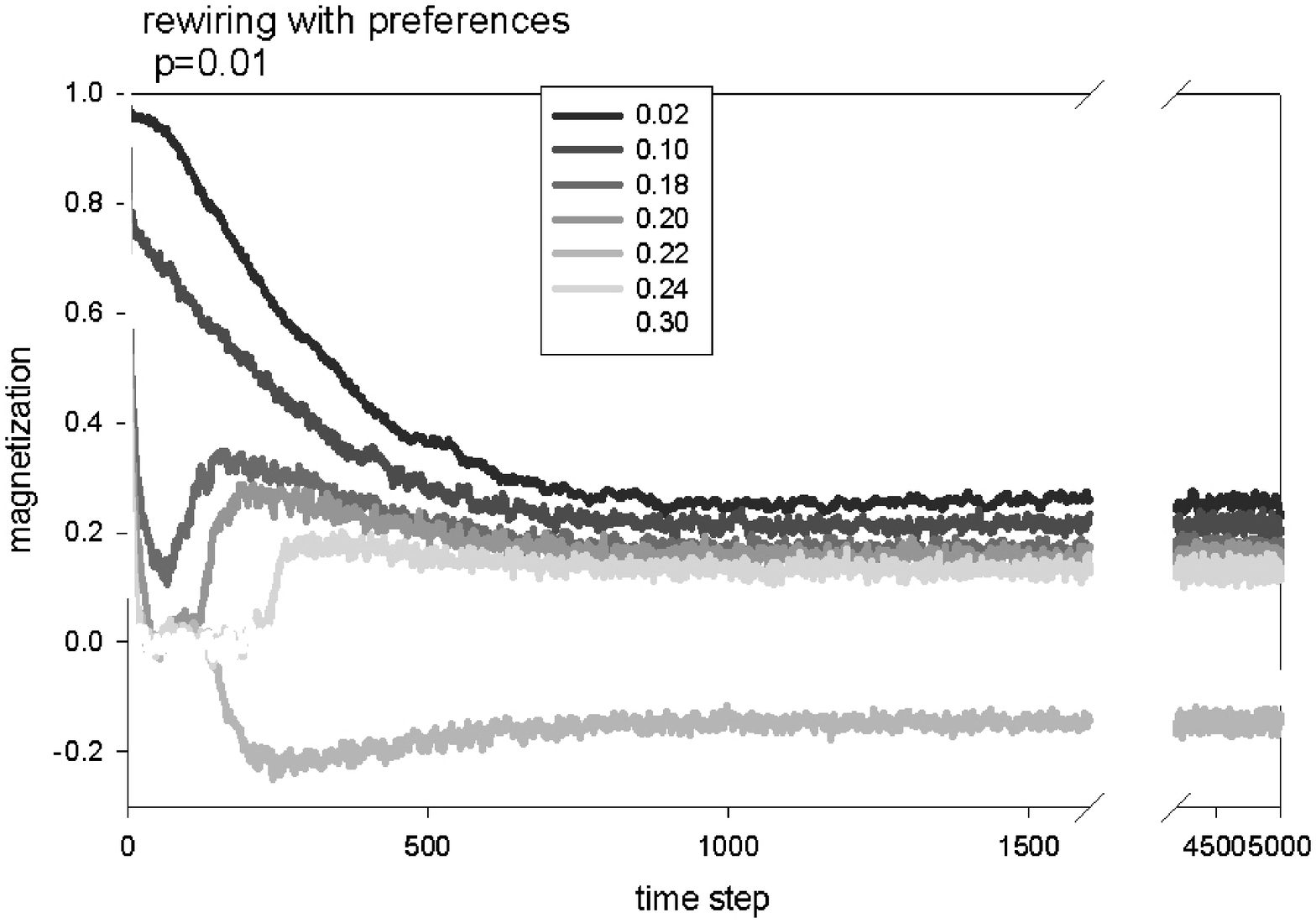}
    \includegraphics[width=0.42\textwidth]{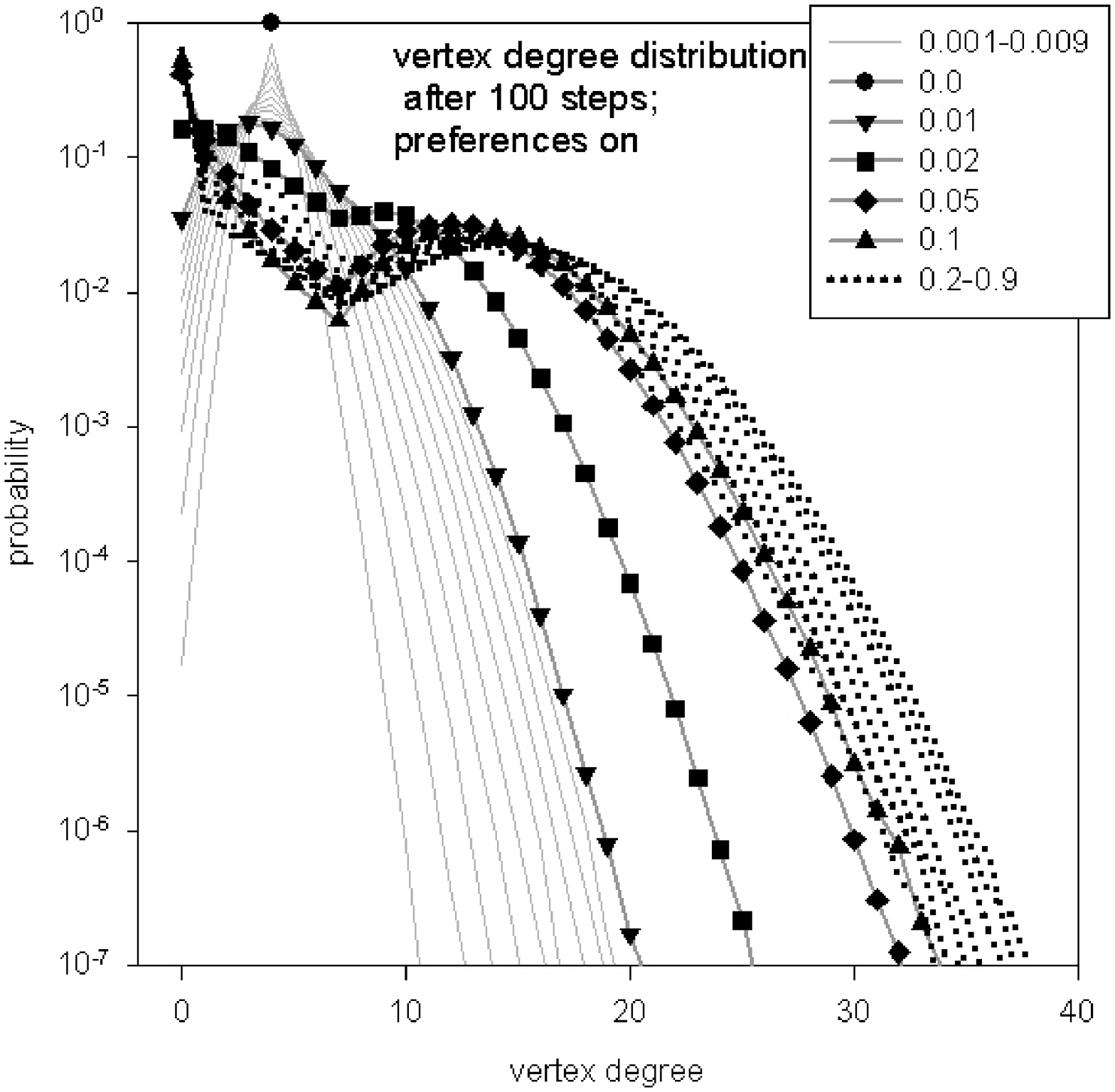}
    \includegraphics[width=0.5\textwidth]{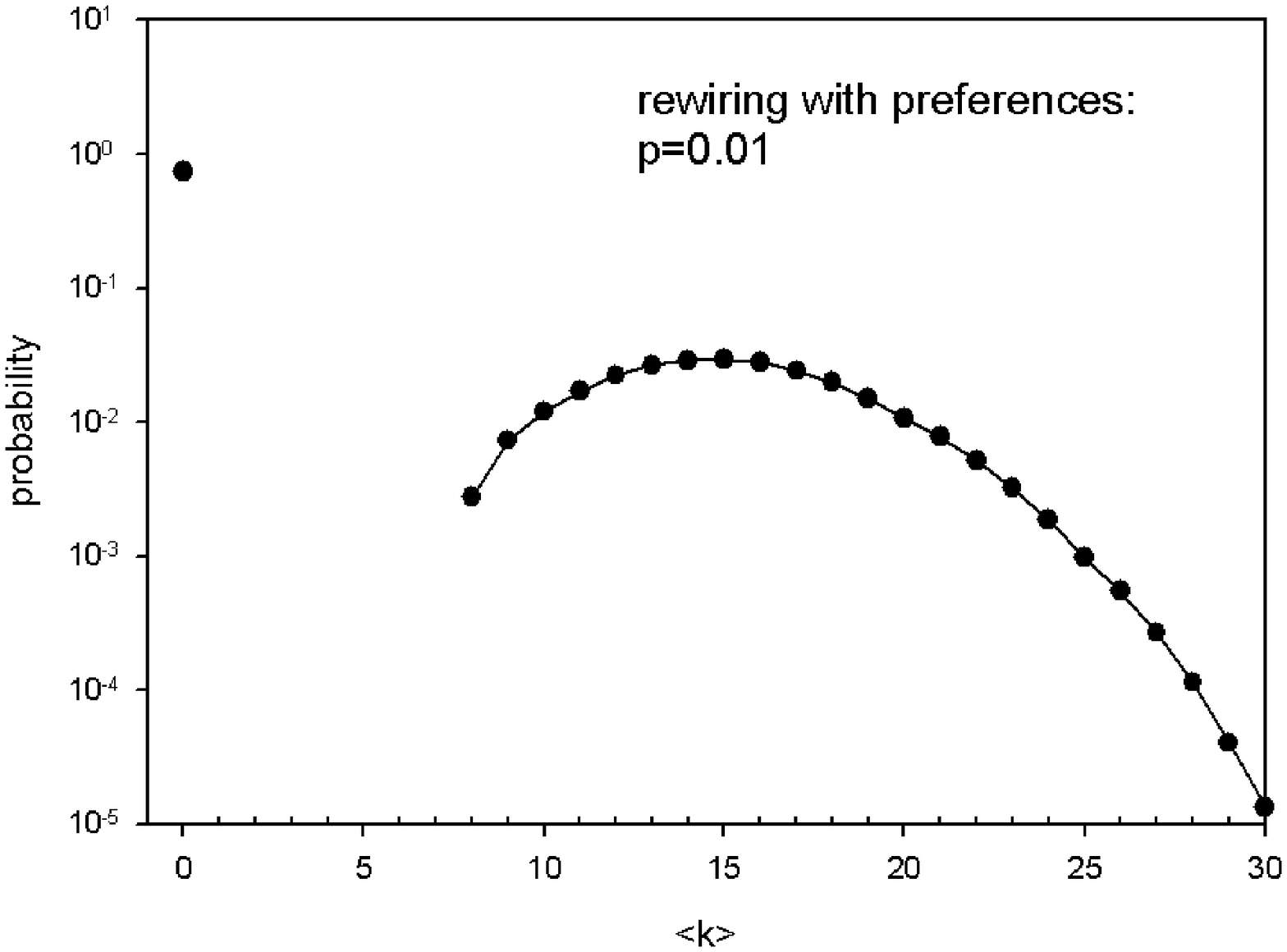}
    \end{center} 
\caption{\newline\small
Typical trajectory of magnetization at different temperature noise $\varepsilon: 0.02,\dots 0.30$ on the network rewired intentionally,  $p=0.01$ (upper panel).  The threshold for preferences is T=8. 
Distribution of vertex degree obtained after 100 time steps (left bottom) and after 10000 steps (right bottom).} 
\label{time_pr}
\end{figure}

\hspace{3.5mm}
If the preferences (b) in edge rewiring are switched on then strong modification to the network appears, see Figure \ref{time_pr}(bottom). The results presented are collected for $T=8$. If the evolution of a network is short (let us assume a hundred step evolution) then the distribution of vertex degree spreads up to 40. For $p$ large the distribution function looks like being composed of two parts --- the exponential part  with rapid decay until the threshold value $T$, and  the  gaussian part  with the mean value slowly moving to  higher degree vertices. When study the phase transition one needs long run observations.  What is  recorded  when the evolution is really long,  the increase in number of isolated vertices is observed. The  distribution presented on Figure \ref{time_pr} (bottom right)is obtained  after 10000 steps.  About 73\% of nodes are isolated vertices. Therefore the magnetization of the stationary state cannot depend on $\varepsilon$, see Figure \ref{time_pr} for a typical trajectory of magnetization.

Hence the phase transition  is not observed --- neither by change in the magnetization value nor by a maximum in susceptibility. The majority rule does not have influence to the state of  isolated nodes. However, the phase transition could be investigated if one  ignores isolated nodes and  concentrates on  magnetic properties of the connected component. Such a solution was proposed  and applied by Herrero \cite{Herrero} in his study of the Ising system on static small-world  network. But on the small-world network only small amount of vertices are 0-degree, compare Figure \ref{stochastic}(bottom panel). Therefore, in the case considered here such approach is unjustified. 

\subsection{Edges evolution with preferences and preserving connectivity demand}
\begin{figure} 
    \begin{center} 
    \includegraphics[width=0.87\textwidth]{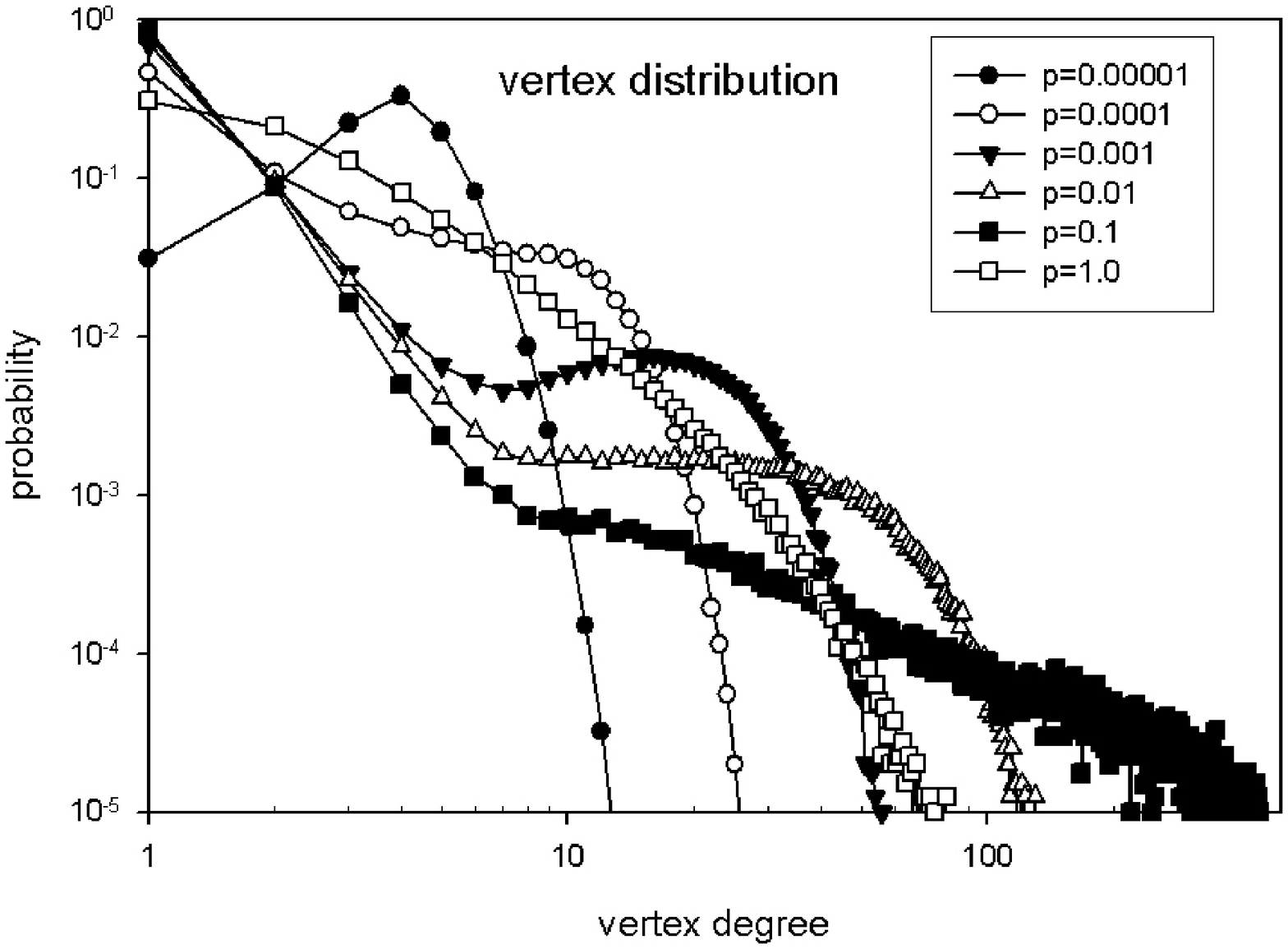}
    \end{center} 
\caption{\newline\small
Distribution of vertex degree in stationary states for different values of $p$, log-log plot.} 
\label{ed_pr_dm}
\end{figure}

\hspace{3.5mm}
To manage with increasing number of isolated nodes the additional condition is introduced --- while rewiring, the graph must stay connected. Thanks to this demand the stationary  networks present properties different from networks considered so far, see Figure \ref{ed_pr_dm} and \cite{Danuta} for details.

Depending on the probability $p$ the distribution of vertex degree shows the transition from an exponential network which is closely related to the  initial regular lattice ($p<10^{-5}$)  to an exponential network where vertex degrees are spread from one to one hundred ($p=1$), by passing through the scale-free networks. Especially, when $p$ is close to $0.1$ the distribution is of power-law type and there are many vertices with degree greater than 500 on the network.

\begin{figure} 
    \begin{center} 
    \includegraphics[width=0.90\textwidth]{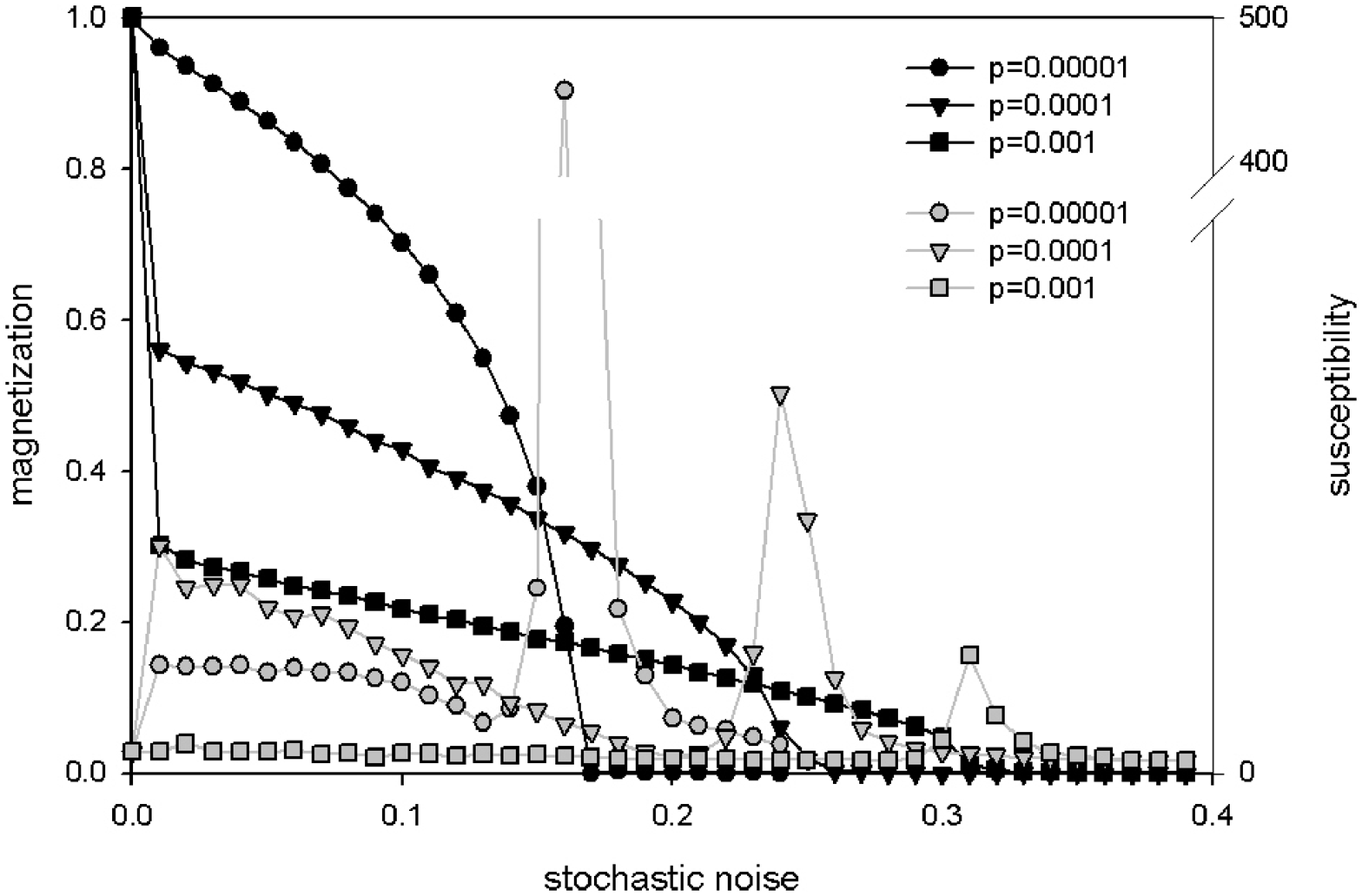}
    \includegraphics[width=0.90\textwidth]{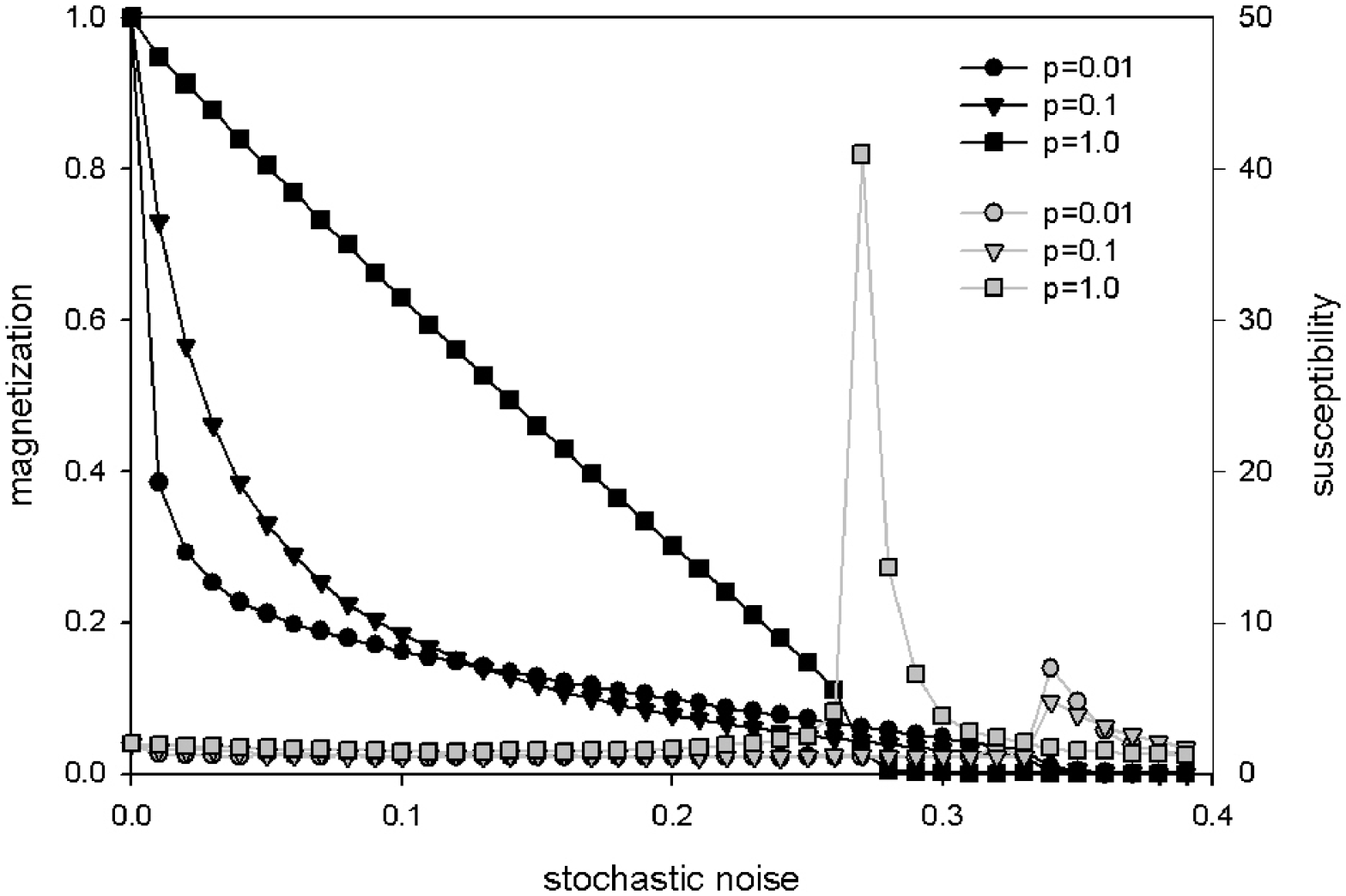}
    \end{center} 
\caption{\newline\small
Magnetization and susceptibility vs temperature noise  $\varepsilon: 0.02,\dots 0.40$ on the network rewired intentionally and with preserving connectivity for   $p=0.00001, 0.0001, 0.001$ - `classic' case (upper panel) and for $p=0.01,0.1, 1.0$ (bottom panel).  The threshold for preferences is $T=8$. } 
\label{magn_pr_dm}
\end{figure} 

The qualitative different magnetic properties are observed while $p$ is changed. In particular, if $p$ is  small, i.e., $p<0.0001$,  then only few edges are rewired at each time step. Therefore the majority rule dynamics dominates over the network evolution  and the magnetization and susceptibility  are similar to those observed when stochastic rewiring  is applied, compare Figure \ref{epsi_crit}(left panel). However,  the magnetization of the stationary system even at low temperature noise is only little different from zero. The permanently modified network amplifies the temperature noise making different temperatures almost indistinguishable. However the rapid change in the susceptibility ensures that the ferromagnetic phase transition takes place. The point of the change $\varepsilon_{crit}$ is located at higher  values of the noise when $p$ increases similarly to the property  observed in case of stochastic rewiring.

For large  $p$, namely $p> 0.01$, the qualitative and quantitative change in the way which system arrives to the stationary state is observed. The stationarity is reached very quickly and the magnetization of the stationary state is higher than the magnetization observed for small $p$. This new feature is related to the fact that at the large $p$  the network is self-organized to the completely different structure from the initial regular lattice. The network organization is as follows: there is a skeleton made of highly connected nodes and plenty of vertices of degree is one or two that are attached to the skeleton nodes. The spin state of any vertex from the skeleton settles the magnetization of plenty other vertices. Therefore, surprisingly, the point of transition $\varepsilon_{crit}$ moves back to low temperature noise, see \ref{magn_pr_dm}(bottom panel)
---  the transition takes place  in the completely different system, namely, on the scale free network, compare \cite{Aleksiejuk}.

\section{Conclusion}
Stationary states of cellular automata with stochastic majority rule and network connections evolving have been investigated by simulations. Three types of rewiring edges were considered. The algorithm of  Watts and Strogatz  was applied at each time step to establish the  network evolving stochastically (the first network). The preference functions were chosen to amplify the role of strongly connected vertices (the second network). The additional demand  to preserve graph connectivity has provided the third way of edge rewiring. Each of the processes listed above yielded the different network:  small-world, scattered nodes and strongly connected component, scale-free network at the properly tuned rewiring, respectively.

The stochastic majority rule applied to these evolving networks has driven  the system  to the ferromagnetic phase transition independently of rewiring intensity only if  the network of first type was examined. A change of transition properties from the regular lattice to the mean-field characteristics was  expected  and was observed by us.  The  huge number of isolated nodes  which were created due to the preferences introduced,  made the thermodynamic investigations meaningless.

Finally, in case of the third  network some properties which are typical for the small-world    transition  were observed but only if the edge rewiring  was of little intensity. However, while pure stochastic evolution of a network results weakly on the total magnetization, the preferences together with the connectivity condition yielded the efficient decrease in the magnetization of stationary state.  The  other transition, called  scale-free ferromagnetic transition, was observed when the parameter $p$ of rewiring was high enough. 

The last  result, namely  emergence the scale-free transition  needs  deep investigations. The observations should be verified in wide region of model parameters to establish relations in quantitative way. On the other hand, some  considerations should  focus on the way in which the small-world transition is transfered into the scale- free transition. In particular, the mechanism in which the hubs (--- widely used name for highly connected nodes) influence the ferromagnetic properties should be clarified. 

Let us close with the following remark. Network evolution rules considered here followed  investigations done in the Ising system  and therefore were far from the basic idea of cellular automata design --- locality. There is a need and necessity to include one more condition to the network rule that any rewired edge can be attached only to the node chosen from a list of neighboring nodes.  Such simulations are run and raw results are promising.

{\bf Acknowledgment\\} 
This work is supported by  Polish Ministry of Scientific Research and Information Technology : 
PB/1472/P03/2003/25.\\

\end{document}